\documentclass[aps,twocolumn,showpacs,floatfix]{revtex4-2}

\usepackage{amssymb}
\usepackage{graphicx}
\usepackage{dcolumn}
\usepackage{bm}
\usepackage{amsmath}
\usepackage{soul,color}
\usepackage{textcomp}
\usepackage[table,xcdraw]{xcolor}
\usepackage{times}
\usepackage{tensor}
\usepackage{braket}
\usepackage{enumitem}

\usepackage[colorlinks,linkcolor=red,citecolor=blue]{hyperref}

\begin{document}

\title{Rapid passage to ordered states in Rydberg atom arrays }
\author{Liang Zeng$^{1}$}
\altaffiliation{These authors contributed equally to this work}
\author{Fei Zhu$^{1}$}
\altaffiliation{These authors contributed equally to this work}
\author{Li Chen$^{1}$}
\email{lchen@sxu.edu.cn}
\author{Heng Shen$^{2}$}
\email{hengshen@nbi.dk}

\affiliation{
$^1${Institute of Theoretical Physics, State Key Laboratory of Quantum Optics Technologies and Devices, Shanxi University, Taiyuan 030006, China} \\
$^2${State Key Laboratory of Quantum Optics Technologies and Devices, Institute of Opto-Electronics, Shanxi University, Taiyuan, Shanxi 030006, China}
}

\begin{abstract}

Given a finite lifetime, a ubiquitous challenge in quantum systems is how to prepare a target state in the shortest possible time. This issue is particularly relevant for Rydberg atom arrays in optical tweezers where the dephasing time is typically restricted to a few microseconds. In this work, we develop a rapid passage to ordered many-body states in a Rydberg atomic chain, which allows the transition of the system to various ordered phases in the phase diagram, such as the Z$_2$-, Z$_3$-, and Z$_4$-ordered antiferromagnetic states. Our scheme ramps the parameter in a "non-adiabatic to quasi-adiabatic to non-adiabatic (NQN)" manner. The NQN configuration significantly reduces the time cost required for the state preparation using entirely adiabatic methods, and is generally appliable to sizable number of atoms. Moreover, we experimentally validate the NQN scheme on the neutral-atom quantum cloud computer Aquila.

\end{abstract}

\maketitle

\section{Introduction}
Ordered states, featuring the regular arrangement of elements over macroscopic scales and long-range correlations, fundamentally constitute our physical world. Their formation is intimately related to the principle of symmetry. According to Landau theory, ordered phases emerge from the spontaneous breaking of a global symmetry of the Hamiltonian, where the system prefers a ground state with lower symmetry. This mechanism is pivotal for understanding critical phenomena \cite{Sachdev2011,Cardy1996} and underpins diverse physical phenomena, ranging from superconductivity \cite{tinkham2004,Tsuei2000}, superfluidity \cite{Leggett1999}, various magnetic orders \cite{Fradkin2013} to liquid crystal phases \cite{Gennes1993}. Ordered states also serve as crucial resources in quantum technology, which are typically used as initial states for exploring many-body dynamics and preparing certain quantum states. For instance, squeezed states, which enable sensing surpassing the standard quantum limit, are commonly prepared via the non-linear evolution of ordered spin coherent states \cite{Gross2012}. Furthermore, the superfluid phase in ultracold atoms within optical lattices, characterized by off-diagonal long-range order \cite{YangCN1962}, offers a fundamental platform for simulating complex condensed matter matters \cite{Endres2011,Amico2008,Gross2017}.

However, it is challenge for artificial quantum systems to efficiently prepare target quantum many-body states with certain orders \cite{Browaeys2020,Cui_2017,XinYuLuo2017,Bason2012,Raimond2001,Malinovskaya2017}. The primary obstacles are the limited lifetime quantum devices. In this sense, it is in highly demand to develop methods for rapidly and reliably preparing target states. Rydberg atom arrays \cite{Bloch2008,Saffman2010, DanielBarredo2016,Balewski2024,Cesa2023,Henriet2020}, renowned for their strong and tunable long-range interactions, have emerged as a leading platform for realizing and manipulating ordered phases. Experiments conventionally start by initializing these arrays with all atoms in the ground state \cite{Bernien2017}, i.e. a disordered ground state $\ket{D}$, and systems are subsequently driven towards desired ordered states through dynamical control. Yet, the dephasing times in Rydberg systems are notably short, often just a few microseconds, limited by factors such as two-photon transition decoherence, atomic thermal motion, intermediate-state scattering and laser noise \cite{Jonathan2023}. This limited coherence window significantly restricts the time available for both state preparation and subsequent quantum simulations or computations, underscoring the critical need for faster preparation techniques specifically tailored for Rydberg atom arrays.

Extensive research has been conducted on state manipulation strategies in Rydberg atom arrays, focusing on both adiabatic \cite{Omran2019,DongXiaoLi2019,Tzortzakakis2022,Schiffer2022} and non-adiabatic approaches \cite{Ostmann_2017,Ji2020,SongPY2024}. Adiabatic schemes typically involve careful scanning of the detuning parameter $\Delta$, facilitating the system's transition from the ground state to antiferromagnetic-like states or crystalline phases \cite{Weimer2010,Pohl2010,Schauß2015}. However, the slow ramping required by adiabatic processes often results in prolonged preparation times. Global chirped laser pulses have also been employed to facilitate rapid adiabatic passage \cite{PhysRevLett.132.153603,Pachniak2021}, dynamically adjusting the system's energy levels to avoid crossing points. On the other hand, non-adiabatic approaches, such as quenches in the Hamiltonian parameters \cite{Schiffer2024,Bluvstein2021,GuardadoSanchez2018}, have proved to be effective with fast transit in the complex energy landscapes of many-body systems. These methods have been successfully leveraged in exploring phenomena such as quantum thermalization and localization \cite{Srednicki1994,annurev2015,Turner.C.J2018,Deutsch_2018}, the emergence of dynamical revivals and quantum scar states \cite{Turner2018,LinChengJu2019,LinChengJu2019,Serbyn2021}, as well as probing critical exponents through the Kibble-Zurek mechanism \cite{Zurek2005,Keesling2019,Ebadi2021}. Moreover, hybrid schemes that combine both adiabatic and non-adiabatic approaches have been developed. For instance, the sweep-quench-sweep (SQS) protocol \cite{lukin2024}, which incorporates a quench between two adiabatic sweeps, can efficiently navigate the system through various ground-state phases in both one-dimensional (1D) and two-dimensional (2D) atom arrays.

In this paper, we concentrate on a 1D Rydberg atom array. Using the Broyden–Fletcher–Goldfarb–Shanno optimization algorithm \cite{BROYDEN1970,Fletcher1970,Goldfarb1970,Shanno1970}, we identify a pathway that enables fast transitions from a disordered ground state to the Neel ordered states that break the Z$_2$, Z$_3$, and Z$_4$ translational symmetries. Our approach achieves these transitions within $\sim$ 2 $\mu s$, outperforming traditional adiabatic methods in terms of efficiency.
We adjust the parameter using a "non-adiabatic to quasi-adiabatic to non-adiabatic" (NQN) hybrid structure, which differs from the previously mentioned SQS scheme—in fact, the structures of the two schemes are inverse. In our scheme, the initial and final non-adiabatic segments rapidly pump the system from the ground state to a certain excited state that smoothly connects to the target state. This configuration also relaxes the stringent conditions typically necessary for adiabatic evolution in the intermediate segment, thereby boosting the overall efficiency of the state preparation process.
Experimentally, we validate the efficacy of the NQN scheme for preparing Z$_2$ states on the quantum cloud computer Aquila.

The rest of this paper is organized as follows: In Sec.~\ref{Model}, we provide a concise overview of the 1D Rydberg atom array system, its equilibrium properties, and the adiabatic state preparation used to achieve ordered states.
Sec.~\ref{passage} formally introduces our rapid passage methodology, detailing the NQN scheme and its operational mechanics. 
In Sec.~\ref{experiment}, we describe the application of the NQN algorithm on the quantum cloud computer Aquila and present the experimental results. 
Finally, a brief summary and discussion are presented in Sec.~\ref{Summary}.

\section{Model and Adiabatic Preparation} \label{Model}

We consider an atomic chain with $N$ atoms arranged in equal lattice spacing $a$. Each atom trapped by an optical tweezer can be treated as a spin-1/2 qubit, where the atomic ground state $\ket{g}$ and the highly excited Rydberg state $\ket{r}$ encode the qubit states $\ket{0}$ and $\ket{1}$, respectively. The dynamics of this 1D Rydberg atom array are governed by the Hamiltonian (setting $\hbar = 1$)
\begin{equation}
    H = \sum_i \left(\frac{\Omega_i}{2} \sigma_i^x - \Delta_i n_i\right) + \sum_{i<j} V_{ij} n_i n_j,
    \label{H}
\end{equation}
where $\Omega_i$ represents the Rabi frequency of the $i$-th atom, $\sigma_i^x = \ket{0_i}\bra{1_i} + \ket{1_i}\bra{0_i}$ describes the laser coupling between the ground and Rydberg states, $\Delta_i$ denotes the detuning of the laser from resonance, and $n_i = \ket{1_i}\bra{1_i}$ is the Rydberg-state occupation number operator for the $i$-th atom. The interaction term $V_{ij} = C_6/|r_i - r_j|^6 = C_6/(|i- j|^6 a^6)$ represents the van der Waals interaction between Rydberg-excited atoms $i$ and $j$, which arises due to the high polarizability of Rydberg states. Here, $C_6$ is the van der Waals coefficient characteristic of the specific Rydberg atomic level involved.

A critical mechanism in this setup is the Rydberg blockade, where the strong interaction between atoms in Rydberg states within a certain radius, known as the blockade radius $R_b$, prevents multiple nearby atoms from being simultaneously excited to the Rydberg state. Consequently, within $R_b$, only one atom can be in the Rydberg state. The blockade radius is quantified by the condition $V_{ij} \geq \Omega_i$, and can be approximated by $R_b = (C_6/\Omega)^{1/6}$.

\begin{figure}[pbt]
	\begin{center}
		\includegraphics[width=.40 \textwidth]{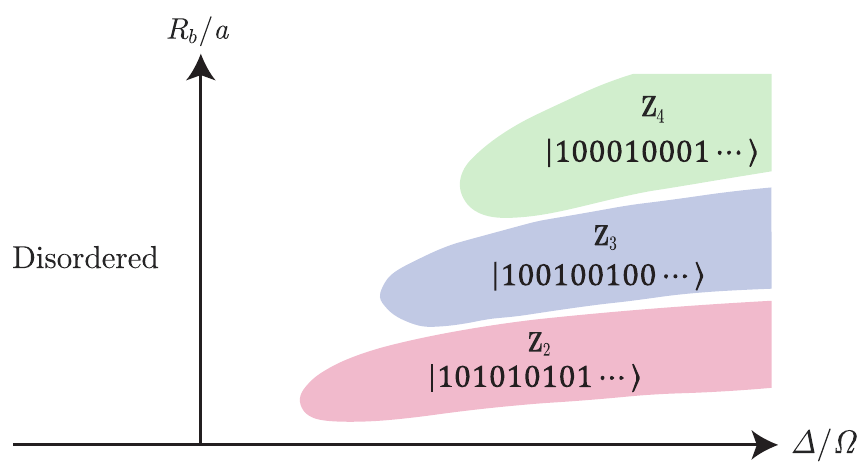}
	\end{center}
		\caption{Phase diagram of 1D Rydberg atom array. $\Delta$ and $a$ respectively denote the detuning and the lattice constant, and $R_b$ represents the blockade radius. Shading areas indicate various ordered phases.}
	\label{pd}
\end{figure}

The equilibrium phase diagram of the system within the $\Delta$-$R_b$ plane has been extensively studied \cite{Bernien2017,Keesling2019}, as shown in Fig.~\ref{pd}. At large negative detunings, the many-body ground state of the system predominantly favors all atoms in their individual atomic ground states, i.e., the disordered ground state $\ket{D} = \prod_{i=1}^N\ket{0_i}$, since Rydberg excitations would increase the total energy. Conversely, at large positive detunings, the system energetically favors maximizing the number of Rydberg excitations, subject to the occupation constraints imposed by the Rydberg blockade effect. The interplay between detuning and blockade radius crucially shapes the ground state configuration.
Specifically, for a blockade radius of $a < R_b < 2a$, where only nearest neighbors are blockaded, the ground state consists of two-fold degenerate $\rm{Z}_2$ antiferromagnetic states, i.e., $\ket{\rm{Z}_2} = \ket{1 0 1 0 \cdots}$ and $\ket{\bar{\rm{Z}}_2} = \ket{0 1 0 1 \cdots}$. These states arise because each excited Rydberg atom blockades its immediate neighbor from also being excited, breaking the $\rm{Z}_2$ translational symmetry. As the blockade radius increases beyond $2a$, encompassing next-nearest neighbors and beyond, more complex ground state configurations emerge. For instance, with $2a < R_b < 3a$, the ground states are $\rm{Z}_3$ ordered antiferromagnetic states, which are three-fold degenerate, exemplified by $\ket{\rm{Z}_3} = \ket{1 0 0 1 0 0 \cdots}$ and others. 

Adiabatic state preparation is commonly used to achieve ordered states. Starting from the disordered ground state $\ket{D}$, the detuning $\Delta$ is scanned from large negative values to positive values to guide the system through a transformation into targeted ordered states. For instance, in Ref.~\cite{Jonathan2023}, the ordered state $\ket{\rm{Z}_2}$ has been adiabatically obtained in about $4 \mu s$ in a Rydberg chain with $N=7$. The implications of the ground-state degeneracy on adiabatic state preparation are significant and hinge critically on whether the total number of particles $N$ is odd or even. Essentially, the adiabatic theorem always necessitates a unique ground state to function effectively. For even $N$, the ordered states are strictly degenerate, which means that the adiabatic theorem is inevitably violated when the $\Delta$ scan reaches the critical point. In contrast, for an odd $N$, there always exists an energy gap $\delta E$ between the ordered states, which allows the adiabatic theorem to remain effective. However, at the critical point, this energy gap decays with increasing $N$, i.e., $\delta E \propto N^{-\gamma}$ with $\gamma$ being a model-dependent exponent. Consequently, for large $N$, the adiabatic preparation requires significantly longer times due to the reduced energy gap. 
The same principles are relevant to other ordered states.

As mentioned earlier, the challenges of adiabatic state preparation are compounded by decoherence and instabilities such as thermal motion and laser noise. For example, in the Rydberg quantum cloud computer Aquila \cite{Jonathan2023}, the dephasing time is restricted to approximately $5.8 \mu s$, setting the maximum operational limit at $4 \mu s$. This constraint significantly hinders the practicality of employing slow adiabatic sweeps, especially in systems hosting a large number of quantum particles $N$. Therefore, there is a pressing need to explore state preparation schemes that transcend traditional adiabatic methods. In fact, fast initial state preparation would not only free up more operational time but also allow for more complex subsequent tasks such as quantum dynamics, computation, and measurement. 

\section{NQN Ramping Scheme} \label{passage}
\subsection{General Passage} \label{passageA}	

Here, we formally introduce our state preparation scheme. Our initial state is the disordered state $\ket{\psi(0)} = \ket{D}$, which is approximately the eigenstate of the system for large negative detuning $\Delta$. Our goal is to find an optimal sweeping path for $\Delta(t)$ such that the system transitions to an ordered state, exemplified here by the $\rm{Z}_2$-ordered state $\ket{\rm{Z}_2}$ with $N \in \text{odd}$. The condition $N \in \text{odd}$ is necessary. As mentioned before, if $N \in \text{even}$, the disordered ground state $\ket{D}$ preserves $\rm{Z}_2$-symmetry, while the ordered state $\ket{\rm{Z}_2}$ breaks $\rm{Z}_2$-symmetry. In quantum dynamics, a unitary evolution cannot spontaneously break the symmetry of a state if the Hamiltonian preserves that symmetry. However, the $\rm{Z}_2$-symmetry of $H$ is explicitly broken for $N \in \text{odd}$. Similar considerations also apply to other ordered states.

The task of obtaining $\Delta(t)$ is essentially a dynamic optimization control problem.
To address this, we fix the Rabi frequency $\Omega = 1 (2\pi)$ MHz as our energy unit, and set the initial detuning $\Delta(0) = -12 \Omega$ and final detuning $\Delta(\tau) = 12 \Omega$, where $\tau$ represents the total state preparation duration. During the interval $t \in [0, \tau]$, we uniformly divide the time into $n_\tau = 8$ segments, during each of which $\Delta(t)$ linearly ramps from $\Delta(\tau_i)$ to $\Delta(\tau_{i+1})$. Note that, although finer time divisions are feasible as long as they exceed the experimental instruments' minimum resolution (e.g., $0.05\mu$s for Aquila), they are not necessary. Our practical simulations have shown that $n_\tau = 8$ can already yield satisfactory results for $\tau \lesssim 2\mu s$, as we will demonstrate below.

We employ the Broyden–Fletcher–Goldfarb–Shanno (BFGS) algorithm to obtain the optimized trajectory $\Delta(t)$. The algorithm adjusts the intermediate control parameters $\Delta(\tau_i)$ for $i \in (2, n_\tau -1)$ by minimizing a function (also called the loss function) 
\begin{equation}
\mathcal{L}[\Delta(t)] = 1 - |\langle \rm{Z}_2 | \psi(\tau) \rangle|^2 = 1- \mathit{F}_{2}(\tau),
\end{equation}
which represents the deviation of the final state from the desired target state, with $F_2(t) = |\langle \rm{Z}_2 | \psi(t) \rangle|^2$ being the fidelity of $\ket{\psi(t)}$ to $\ket{\rm{Z}_2}$.
Technically, the BFGS algorithm approximates the Hessian matrix of the loss function $\mathcal{L}$, updating this approximation iteratively based on gradient evaluations. The algorithm is known for its robustness and efficiency in handling complex optimization landscapes without requiring the actual Hessian but using an approximation that evolves with each iteration.
Although BFGS is a local optimization algorithm, we enhance its capability to approximate global optimization by initializing the algorithm at multiple points in the parameter space. For this study, we typically perform 50 such initializations for each set of system parameters. 

\begin{figure}[t]
	\begin{center}
		\includegraphics[width=.49 \textwidth]{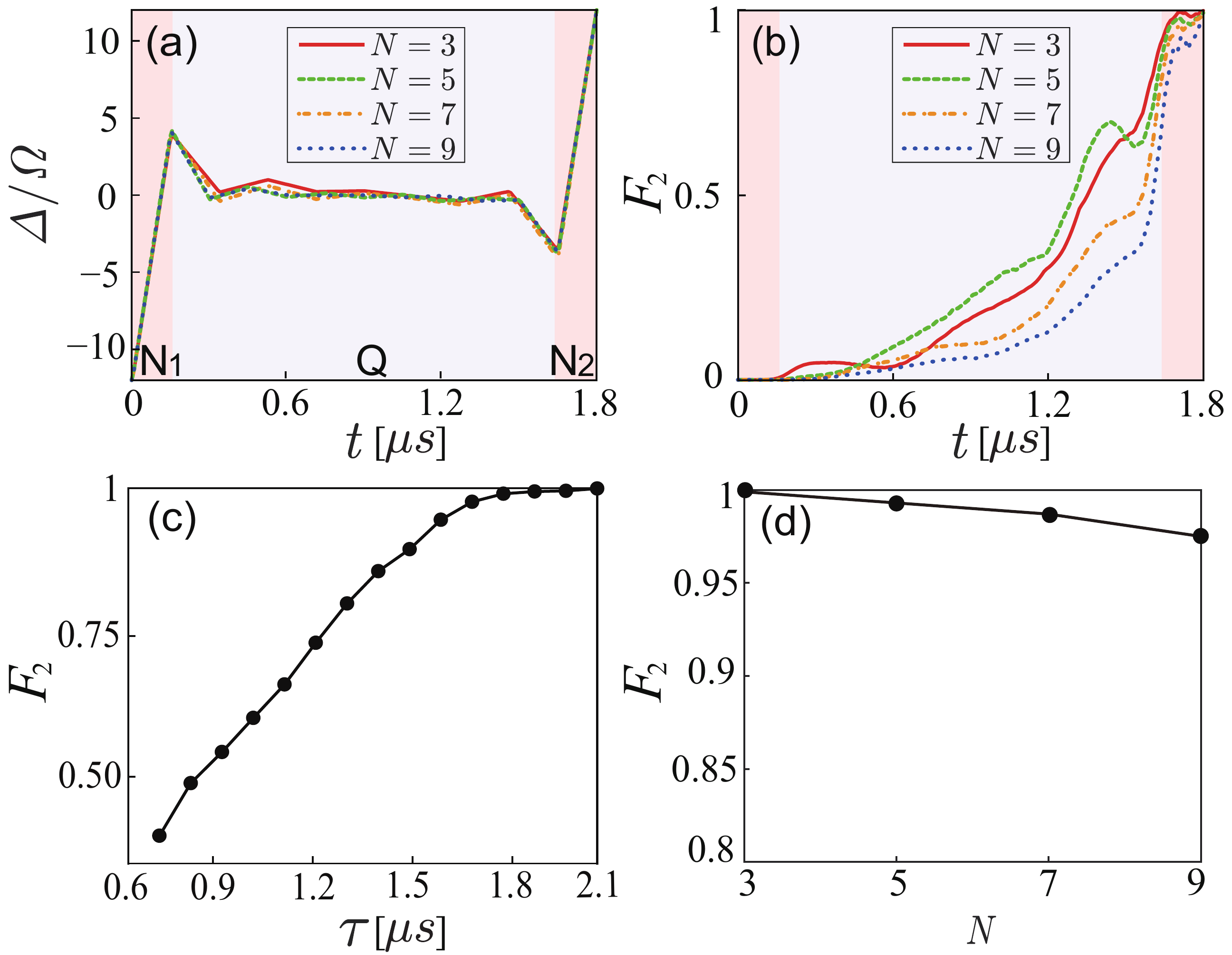}
	\end{center}
		\caption{Preparation of $\ket{\rm{Z}_2}$ state with the NQN scheme, with $\Omega = 1$ MHz being fixed and serving as the energy unit. (a) Temporal series of $\Delta(t)$, with various line styles indicating different particle numbers $N$. The colored shadings indicate the N$_1$, Q, and N$_2$ segments of the NQN scheme, respectively. (b) Corresponding dynamics of the fidelity $F_2(t)$ to $\ket{\rm{Z}_2}$ state. (c) The final-state fidelity $F_2(\tau)$ as a function of the total preparation time $\tau$ with fixed $N=7$. (d) $F_2(\tau)$ as a function of $N$ with fixed $\tau = 1.8 \mu s$.}
	\label{fig2}
\end{figure}

Using the optimized trajectory $\Delta(t)$ obtained through the BFGS algorithm, we can achieve the final state with high fidelity to $\ket{\rm{Z}_2}$. In Fig.~\ref{fig2}(a) and (b), we respectively present the optimized detuning $\Delta(t)$ and the corresponding fidelity $F_2(t)$ for a fixed $\tau = 1.8 \mu s$, where various line styles indicate different particle numbers $N$. The results clearly show that $F_2(\tau) \approx 1$, i.e., the fidelity of the final state is close to $100\%$. This outcome underscores the efficiency of our approach compared to traditional adiabatic methods.

More importantly, as indicated in Fig.~\ref{fig2}(a), the curve $\Delta(t)$ is approximately symmetric about $t = \tau/2$, featuring rapid sweeps of $\Delta$ at the beginning and end (marked by red shadings), with a slow adjustment in the middle (marked by blue shading). We term the structure of $\Delta(t)$ as the "non-adiabatic to quasi-adiabatic to non-adiabatic" scheme, i.e., the NQN scheme. The "NQN" respectively denote: N$_1$ segment with non-adiabatic forward sweep; Q segment exhibiting a quasi-adiabatic backward sweep; N$_2$ segment with a non-adiabatic forward sweep. Fig.~\ref{fig2}(a) also clearly shows that the NQN scheme exhibits considerable robustness across various values of $N$. Since the BFGS algorithm does not inherently guarantee a unique optimal solution, the ability of the scheme to apply across different $N$ suggests a possible underlying universality in the dynamics.

In Figs.~\ref{fig2}(c) and (d), we further numerically explore the final-state fidelity $F_2(\tau)$ as a function of the total preparation time $\tau$ and the number of particles $N$. Specifically, in Fig.~\ref{fig2}(c), we fix $N=7$ and vary $\tau$. Here, $F_2(\tau) \approx 1$ for $\tau > \tau_0$ with $\tau_0 = 1.7 \mu s$; however, $F_2(\tau)$ begins to decline when $\tau < \tau_0$. The optimal time $\tau_0$ depends on the particle number $N$. 
In Fig.~\ref{fig2}(d), we show the fidelity $F_2$ by fixing $\tau = 1.8 \mu s$ and varying $N$. It is observed that the fidelity slightly decreases and approximately linearly in $N$, with $F_2(\tau) \gtrsim 96\%$ for $N \lesssim 10$, demonstrating robustness of our scheme in small to medium-sized systems. 

\begin{figure}[t]
	\begin{center}
		\includegraphics[width=.49 \textwidth]{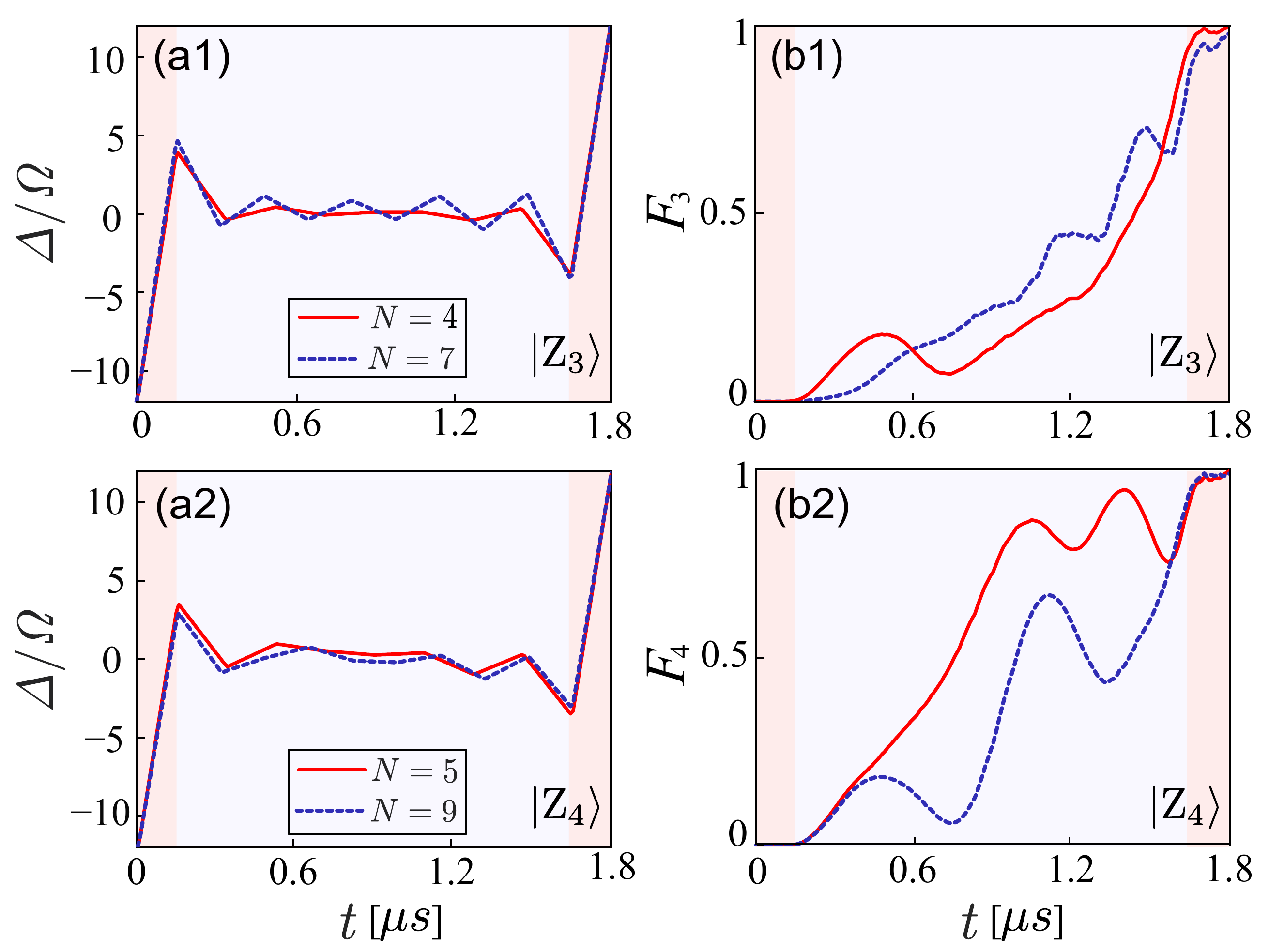}
	\end{center}
		\caption{Preparation of $\ket{\rm{Z}_3}$ [(a1) and (b1)] and $\ket{\rm{Z}_4}$ [(a2) and (b2)] states using the NQN scheme. The left panels (a) display the optimized sweeping curves $\Delta(t)$; the right panels (b) show the dynamics of the fidelities $F_3$ and $F_4$. We take $a \approx 0.36 R_b$ and $a \approx 0.29 R_b$ for the two state-preparation tasks, respectively.}
	\label{fig3}
\end{figure}

Furthermore, we also demonstrate the applicability of the NQN scheme to other ordered states in the phase diagram. The efficacy of the algorithm for achieving Z$_3$-ordered states $\ket{\rm{Z}_3}$ and Z$_4$-ordered states $\ket{\rm{Z}_4}$ is showcased in the first row [Figs.~\ref{fig3}(a1) and (b1)] and the second row [Figs.~\ref{fig3}(a2) and (b2)], respectively. Again, the left panels (a) display the optimized sweeping curves $\Delta(t)$ for various particle numbers $N$, whereas the right panels (b) show the dynamics of the fidelities $F_3(t)$ and $F_4(t)$, where $F_3(t) = |\langle \rm{Z}_3 | \psi(t) \rangle|^2$ and $F_4(t) = |\langle \rm{Z}_4 | \psi(t) \rangle|^2$.
In the calculation, we fix $\tau = 1.8 \mu s$ and take $a \approx 0.36 R_b$ and $a \approx 0.29 R_b$ for the preparations of $\ket{\rm{Z}_3}$ and $\ket{\rm{Z}_4}$ states, respectively. 
The figures clearly indicate that the three-stage NQN structure is effectively maintained across both tasks, with the final-state fidelities $F_3(\tau)$ and $F_4(\tau)$ closely approaching 100\%.
The only difference is that the Q segment exhibits some specific variations as $N$ changes. These results demonstrate the generality of the NQN scheme for preparing other ordered states within the phase diagram shown in Fig.~\ref{pd}.

\subsection{Mechanism} \label{passageB}

Now, we discuss the mechanisms underlying the NQN structure, again using the preparation of the $ \ket{\rm{Z}_2} $ state (shown in Fig.~\ref{fig2}) as an illustration. We define the \textit{non-adiabatic basis} of the system as the product states of local bare states, labeled by Roman numerals, i.e., $\ket{\rm{I}} = \ket{0000\cdots}$, $\ket{\rm{II}} = \ket{1000\cdots}$, etc.; and define the \textit{adiabatic basis} as the instantaneous eigenstates of the Hamiltonian $H(t)$, denoted as $\ket{E_n}$ with $n = 1,2,\ldots$ being integers.

In short, the NQN scheme encompasses the following processes:
(1) The rapid sweep of $\Delta$ in the N$_1$ segment pumps the initial state $\ket{\psi(0)}$ into a certain high-energy eigenstate $\ket{E_m}$. (2) The backward sweep in the Q segment transitions the $\ket{E_m}$ into the targeted ordered state $\ket{\rm{Z}_2}$. (3) The final sweep in the N$_2$ segment quickly transitions the $\ket{\rm{Z}_2}$ state into the system's ground state, during which the constructive phase interference eliminates contributions from all non-zero components, leading to the final fidelity close to 100\%. Let us elaborate on these processes in detail.

\begin{figure}[t]
	\begin{center}
		\includegraphics[width=.49 \textwidth]{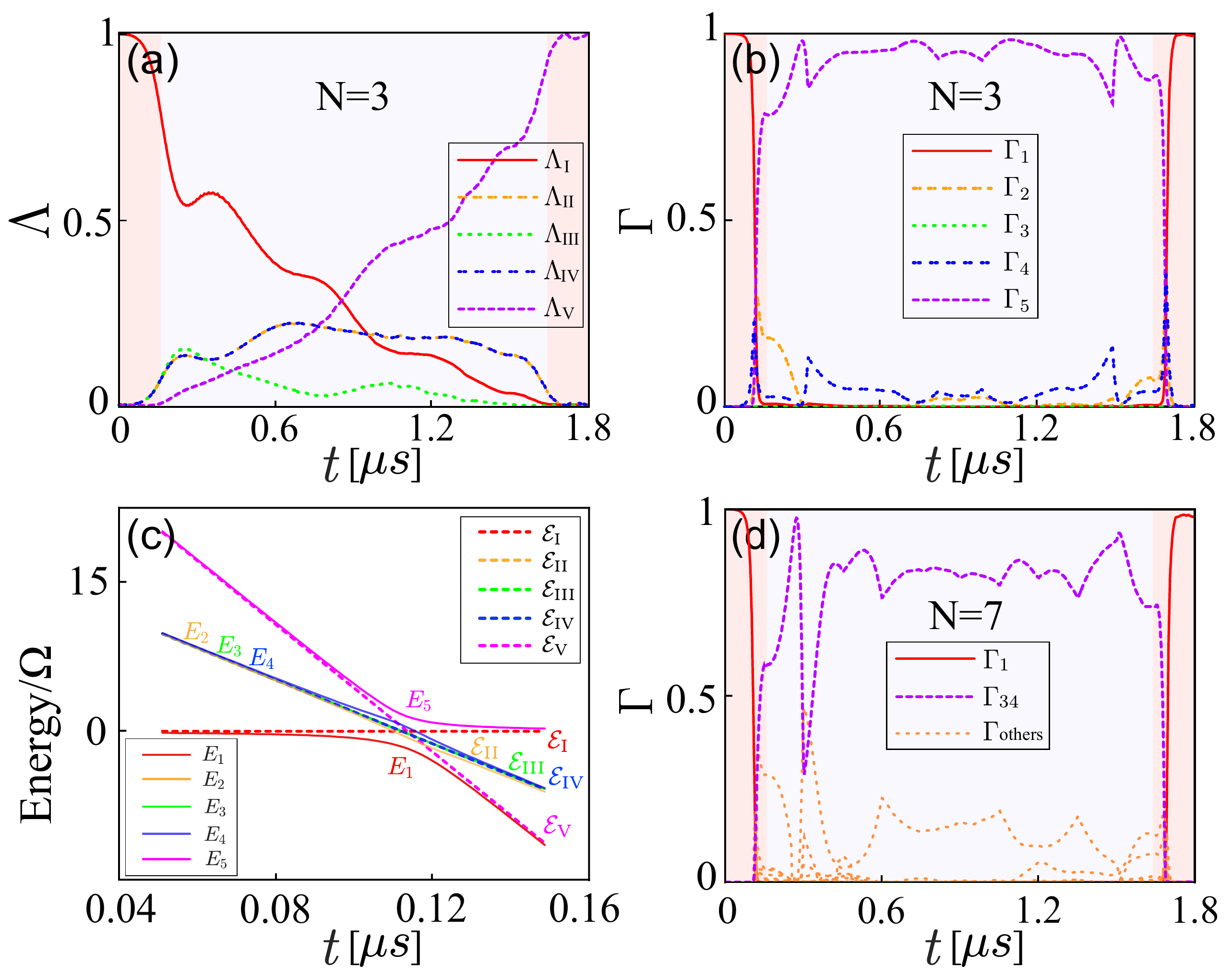}
	\end{center}
		\caption{(a)-(c) Results for the preparation of the $\ket{\rm{Z}_2}$ state with $N=3$. (a) Dynamics of the projection probability $\Lambda$ onto the non-adiabatic basis $\ket{n}$; (b) Dynamics of the projection probability $\Gamma$ onto the adiabatic basis $\ket{E_n}$; (c) Variations of the non-adiabatic-basis energies $\mathcal{E}_n$ and the adiabatic-basis eigenenergies $E_n$ in the N$_1$ segment. (d) Dynamics of $\Gamma$ for the case of $N=7$.}
	\label{fig4}
\end{figure}

Since the NQN scheme works well for sizable $N$, we consider the simplest case of $N=3$, i.e., a three-atom system. We find that the most contributions to the dynamics come from the lowest five non-adiabatic basis states, which are:
\begin{equation}
\left\{\begin{aligned}
\ket{\rm{I}} = & \ket{000}, \ \ \mathcal{E}_{\rm{I}} = 0 \\
 \ket{\rm{II}} = & \ket{100}, \ \ \mathcal{E}_{\rm{II}} = -\Delta(t) \\
 \ket{\rm{III}} = & \ket{010}, \ \ \mathcal{E}_{\rm{III}} = -\Delta(t) \\
 \ket{\rm{IV}} = & \ket{001}, \ \ \mathcal{E}_{\rm{IV}} = -\Delta(t) \\
\ket{\rm{V}} = & \ket{101}, \ \ \mathcal{E}_{\rm{V}} = -2 \Delta(t) + \frac{V}{64}
\end{aligned}\right., \\
\end{equation}
where $\mathcal{E}_{n=\{\rm{I}\cdots\rm{V}\}} = \bra{n}H\ket{n}$ represents the energy of the non-adiabatic basis. Notably, the disordered state $\ket{\rm{I}}$ serves as our initial state, i.e., $\ket{\rm{I}} = \ket{D} = \ket{\psi(0)}$; whereas the antiferromagnetic state $\ket{\rm{V}}$ corresponds to the desired target ordered state, i.e., $\ket{\rm{V}} = \ket{\rm{Z}_2}$.
In such a convention, the Hamiltonian [Eq.~(\ref{H})] can be re-expressed in the non-adiabatic bases as
\begin{equation}
H(t) = 
\begin{pmatrix}
0 & \frac{\Omega}{2} & \frac{\Omega}{2} & \frac{\Omega}{2} & 0 \\
\frac{\Omega}{2} & -\Delta(t) & 0 & 0 & \frac{\Omega}{2} \\
\frac{\Omega}{2} & 0 & -\Delta(t) & 0 & 0 \\
\frac{\Omega}{2} & 0 & 0 & -\Delta(t) & \frac{\Omega}{2} \\
0 & \frac{\Omega}{2} & 0 & \frac{\Omega}{2} & -2 \Delta(t) + \frac{V}{64}
\end{pmatrix}.
\end{equation}
On the other hand, the adiabatic basis $\ket{E_n}$ and the instantaneous eigenenergies $E_n$ can be obtained through numerical diagonalization of $H(t)$. We define the projection probability of $\ket{\psi(t)}$ onto the non-adiabatic basis as 
\begin{equation}
\Lambda_{n=\{\rm{I}\cdots\rm{V}\}} = |\langle n|\psi(t)\rangle|^2,
\end{equation}
and the projection probability of \(\ket{\psi(t)}\) onto the adiabatic basis as
\begin{equation}
\Gamma_{n=\{1\cdots5\}} = |\langle E_n|\psi(t)\rangle|^2.
\end{equation}
Fig.~\ref{fig4}(a) and (b) respectively display the projection probabilities $\Lambda_{n}$ and $\Gamma_{n}$ as functions of $t$ during the preparation of $\ket{\rm{Z}_2}$.

\begin{itemize}[leftmargin=12pt]
\item \textit{N$_1$ segment:} $\Delta$ is swept forward from $-12 \Omega$ to about $3 \Omega$. The non-adiabatic basis state $\ket{\rm{I}}$, which is also the initial state, encounters level crossings with other excited states at $\Delta = 0$, $V/128$, and $V/64$. 
Due to the rapid sweep, these crossing points approximately coincide at $t = t_0 \approx 0.1 \mu s$ on the time axis, as depicted by the dashed lines in Fig.~\ref{fig4}(c). In this figure, the dashed lines represent the energies of the non-adiabatic basis states $\mathcal{E}_n$, while the solid lines represent the adiabatic basis eigenenergies $E_n$, which are always energetically gapped due to the finite $\Omega$. According to Landau-Zener theory, the transition probability between non-adiabatic states at crossing points is given by
\begin{equation}
P = e^{-2\pi {\Omega^2}/{\dot{\Delta}}},
\label{P}	
\end{equation}
where $\dot{\Delta}$ represents the sweep rate. For a large $\dot{\Delta}$, as in the current case, $P$ is close to 1, meaning that the state remains close to the initial state. The result in Fig.~\ref{fig4}(a) shows that the projection probability $\Lambda_\text{I}$ is approximately 80\%, aligning well our theoretical prediction $P(t_0) \approx 0.82$ using Eq.~(\ref{P}). Also shown in Fig.~\ref{fig4}(a), there exist minor contributions from other non-adiabatic states, e.g.,  $\ket{\rm{II}}$-$\ket{\rm{VI}}$, which however will be interfered out in the N$_2$ segment.
On the other hand, from the perspective of the adiabatic basis shown in Fig.~\ref{fig4}(b), due to the severe breakdown of the adiabatic theorem, $\ket{\psi(t)}$ is rapidly pumped from the adiabatic ground state $\ket{E_1}$ to the excited state $\ket{E_5}$.

\item \textit{Q segment:} 
$\Delta$ is slowly swept backward from $3 \Omega$ to $-3 \Omega$. This segment demonstrates quasi-adiabatic behavior, as evidenced by $\Gamma_5$ maintaining a high value, approximately 0.9, as shown in Fig.~\ref{fig4}(b).
In the non-adiabatic picture [Fig.~\ref{fig4}(a)], the state $\ket{\psi} \approx \ket{E_5}$ gradually transitions from $\ket{\rm{I}}$ to the target state $\ket{\rm{V}} = \ket{\rm{Z}_2}$. At the end of this segment, there remains a small contribution from other modes $\ket{\rm{II}}$-$\ket{\rm{IV}}$. The phases of these components have been appropriately adjusted during the evolution such that they can be interfered out in the next segment.

\item \textit{N$_2$ segment:} $\Delta$ rapidly sweeps forward from $-3 \Omega$ to $12 \Omega$. This action resembles the inverse process of the N$_1$ segment: near level-crossing points ($\Delta = 0$, $V/128$, and $V/64$), the Landau-Zener transition with $P \approx 1$ ensures $\ket{\psi(t)} \approx \ket{\rm{V}}$. In this process, the impure non-adiabatic modes $\ket{\rm{II}}$-$\ket{\rm{IV}}$ are coherently eliminated, bringing the final state fidelity close to 100\%. Viewed in the adiabatic basis [Fig.~\ref{fig4}(b)], $\ket{E_5}$ and $\ket{E_1}$ exchange, allowing $\ket{\psi(t)}$ to return to the system's ground state $\ket{E_1}$.
\end{itemize}

\begin{figure}[t]
	\begin{center}
		\includegraphics[width=.49 \textwidth]{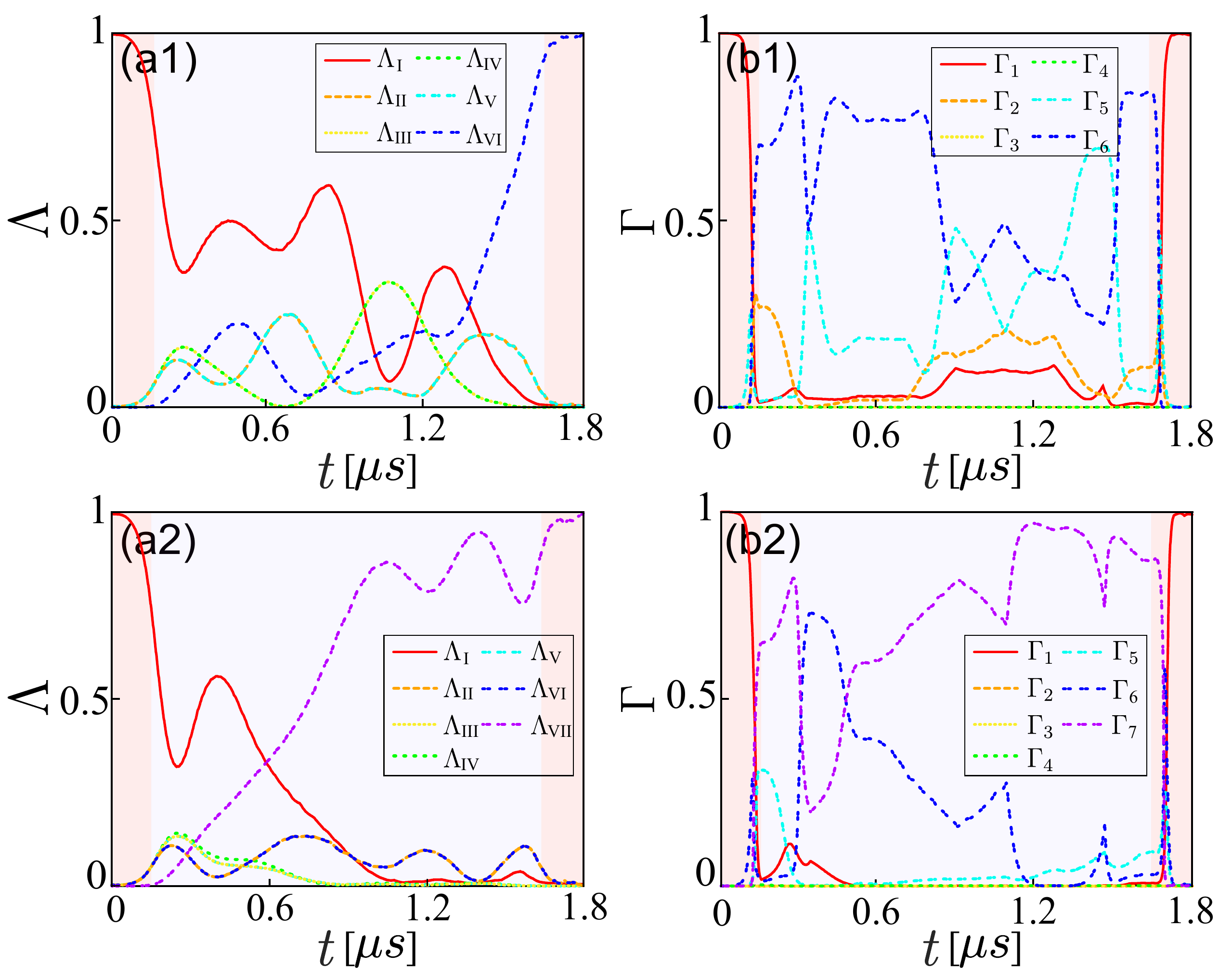}
	\end{center}
		\caption{Dynamics of $\Lambda$ and $\Gamma$ for the preparation of the $\ket{\rm{Z}_3}$ state [(a1) and (b1)] and the $\ket{\rm{Z}_4}$ state [(a2) and (b2)]. $N=4$ and $N=5$ are considered for the tasks regrading $\ket{\rm{Z}_3}$ and $\ket{\rm{Z}_4}$, respectively.}
	\label{fig5}
\end{figure}

Now, we can conduct a comprehensive comparison between the NQN scheme and traditional adiabatic schemes. Firstly, traditional schemes rely on ground state adiabatic transitions, as mentioned in Sec.~\ref{Model}, where the adiabatic theorem must be strictly adhered to, meaning the Landau-Zener transition probability should consistently satisfy $P \approx 0$ during the entire evolution. Therefore, even in regimes far from level crossings, $\Delta$ cannot be swept too rapidly since a large $\dot{\Delta}$ would inevitably generate excitations in the adiabatic basis. In contrast, the N$_1$ segment of the NQN scheme requires the quantum state to rapidly traverse through the level crossings to achieve adiabatic level inversion, i.e., $P \approx 1$, thus $\dot{\Delta}$ in this segment is unrestricted. Secondly, although the Q segment of the NQN scheme resembles adiabatic evolution with the goal of smoothly transitioning the adiabatic state $\ket{E_m}$ from an initial disordered state $\ket{D}$ to the targeted ordered state $\ket{\rm{Z}}$, the condition for implementing this transition is much more relaxed than that for strict adiabatic evolution. Thanks to the coherent elimination in the N$_2$ segment, the NQN scheme exhibits a high tolerance for excitations of adiabatic modes, allowing the Q segment to be executed faster than a purely adiabatic approach. Consequently, these two factors determine the superior efficiency of the NQN scheme compared to the traditional adiabatic approach.

While the discussion above focused on the preparation of the $\ket{\rm{Z}_2}$ state with $N = 3$, the physics applies well to systems with $N > 3$. The only difference lies in the post-transition state $\ket{E_m}$ of the N$_1$ segment: for larger $N$, $m$ will have a value greater than 5. In that case, the trajectory of the N$_1$ segment needs slight adjustment [see $\Delta(t)$ in Fig.~\ref{fig2}(a) with larger $N$] to ensure that the initial state is optimally pumped into an appropriate high-energy eigenstate $\ket{E_m}$ which can then be smoothly transitioned into the target state during the Q segment. 
For example, for $N=5$ and $N=7$, we have $m = 13$ and $m = 34$, respectively. The dynamics of $\Gamma_n(t)$ for $N=7$ are exemplified in Fig.~\ref{fig4}(d), where the dominant contribution of $\Gamma_{34}$ is particularly evident.

We additionally note that the discussion on the preparation of $\ket{\rm{Z}_2}$ is also applicable to the preparation of other ordered states. In Figs.~\ref{fig5}(a) and (b), we show the results for the preparation of $\ket{\rm{Z}_3}$ state with $N=4$ and $\ket{\rm{Z}_4}$ state with $N=5$, respectively. Again, panels (a1) and (a2) represent the projection probabilities $\Lambda$; panels (b1) and (b2) represent the projection probabilities $\Gamma$. 
For the task of preparing $\ket{\rm{Z}_3}$, the non-adiabatic basis are defined as
\begin{equation}
\begin{aligned}
\ket{\text{I}}  &=  \ket{0000}, \ \ \ 
\ket{\text{II}}  =  \ket{1000}, \\
\ket{\text{III}}  &=  \ket{0100},\ \ \ 
\ket{\text{IV}}  =  \ket{0010}, \\
\ket{\text{V}}  &=  \ket{0001},\ \ \ 
\ket{\text{VI}}  =  \ket{1001}; \\
\end{aligned}
\end{equation}
whereas for the preparation of $\ket{\rm{Z}_4}$, the non-adiabatic are defined as
\begin{equation}
\begin{aligned}
\ket{\text{I}}  &=  \ket{00000}, \ \ \ 
\ket{\text{II}}  =  \ket{10000}, \\
\ket{\text{III}}  &=  \ket{01000}, \ \ \ 
\ket{\text{IV}}  =  \ket{00100}, \\
\ket{\text{V}}  &=  \ket{00010}, \ \ \ 
\ket{\text{VI}}  =  \ket{00001}, \\
\ket{\text{VII}}  &=  \ket{10001}.
\end{aligned}
\end{equation}
From the figures, it is evident that the physical outcomes qualitatively align with those observed in the preparation of the $\ket{\rm{Z}_2}$ state. Quantitatively, $\Lambda(t)$ and $\Gamma(t)$ exhibit more fluctuations during the Q segment.

\begin{figure*}[t]
	\begin{center}
		\includegraphics[width=.95 \textwidth]{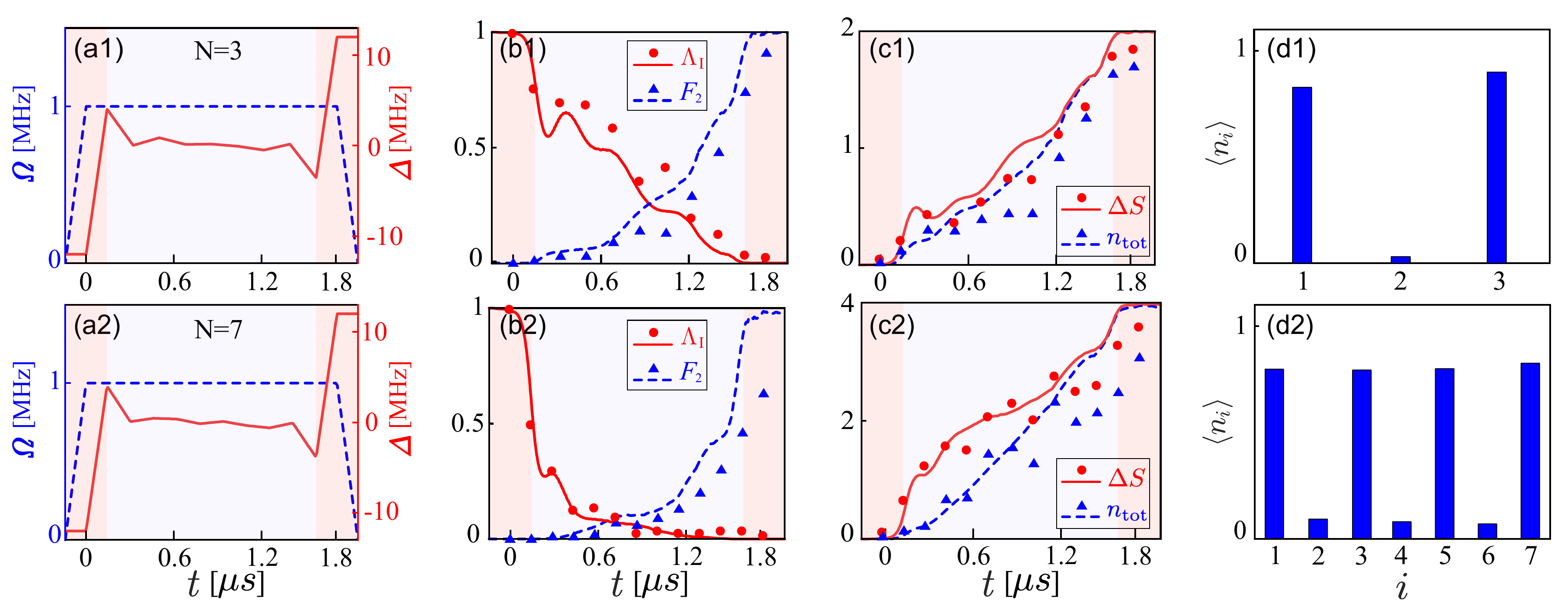}
	\end{center}
		\caption{Experimental results for the preparation of the $\ket{\rm{Z}_2}$ state with $N=3$ (the upper row) and $N=7$ (the lower row) on quantum computer Aquila. (a) Control curves for the system's parameters $\Delta$ (solid line) and $\Omega$ (dashed line). (b) Dynamics of the projection probabilities $\Lambda_\text{I} = |\langle D| \psi(t)\rangle|^2$ and $F_2 = |\langle \rm{Z}_2| \psi(t)\rangle|^2$ with respect to the initial state and the target ordered state. Discrete markers represent experimental data, and the solid curves are theoretical predictions. (c) Dynamics of the antiferromagnetic order parameter $\Delta S$ and the total Rydberg population $n_{\rm{tot}}$. (d) Local Rydberg occupations $\langle n_i \rangle$ of the final state $\ket{\psi(\tau)}$, with $\tau = 1.8\mu s$.}
	\label{fig6}
\end{figure*}

\section{Experimental Validation} \label{experiment}

We experimentally validate our theoretical proposal on the state-of-the-art quantum cloud computer, Aquila, developed by QuEra \cite{Jonathan2023}. Aquila is a neutral-atom quantum computer that leverages $^{87}$Rb atoms, trapped and cooled using optical tweezers. Each atom can be individually encoded as a qubit in its electronic states. The Rydberg states are excited using a two-photon transition, where the energy difference between the ground and Rydberg states is mediated by lasers with wavelengths of 420 nm and 1013 nm. Aquila specifically uses the $\ket{g} = |5S_{1/2}\rangle$ ground state and the $\ket{r} = |70S_{1/2}\rangle$ Rydberg state, characterized by the van der Waals interaction coefficient $C_6 \approx 862,690 (2\pi) \text{ MHz }\mu \text{m}^6$ and blockade radius $R_b \approx 9.76 \mu$m for $\Omega = 1 (2\pi) \text{ MHz}$.
One limitation of the Aquila setup is the minimum interatomic distance being set at $a = 4 \mu\text{m}$. This spacing constraint only allows us to test the NQN scheme for the preparation of the $\rm{Z}_2$-ordered state. For the preparation of $\ket{\rm{Z}_3}$ and $\ket{\rm{Z}_4}$ states, a smaller interatomic distance $a < 3.9 \mu$m is required. As mentioned in Sec.~\ref{Model}, the typical dephasing time of the Rydberg states in this setup is approximately $6 \mu\text{s}$.

The experimental results of implementing the NQN scheme to achieve the $\ket{\rm{Z}_2}$ state on Aquila are illustrated in Fig.~\ref{fig6}, where the first and second rows display the outcomes for $N=3$ and $N=7$, respectively. Specifically, panels (a) demonstrate the experimental control of $\Omega$ (dashed curve) and $\Delta$ (solid curve). The $\Delta$ curves correspond to the NQN curves shown in Fig.~\ref{fig2}(a), modified to include $0.15\mu\text{s}$ idle periods at both the beginning and end. These idle intervals are included to rapidly turn on and off the Rabi frequency $\Omega$, with the latter facilitating measurement operations since $\Omega$ needs to be switched off before measurement \cite{Jonathan2023}. For each evolution, the system undergoes 100 shots to gather statistics on the occupation probabilities $\Lambda_n$ of the measurable non-adiabatic basis states and the local Rydberg population distribution $\langle n_i \rangle$ across the atomic array.

Figure~\ref{fig6}(b) displays the return probability $\Lambda_{\rm{I}}$ of $\ket{\psi(t)}$ to the initial disordered ground state $\ket{D}$, and the dynamical fidelity $F_2(t)$ of $\ket{\psi(t)}$ to the target ordered state $\ket{\rm{Z}_2}$. Smooth curves represent theoretical predictions, while discrete points show experimental data. Fig.~\ref{fig6}(c) shows the evolution of the antiferromagnetic order parameter
\begin{equation}
\Delta S  = \sum_{i \in \text{odd}} \langle n_i \rangle - \sum_{i \in \text{even}} \langle n_{i} \rangle,
\end{equation}
and the total Rydberg population
\begin{equation}
n_{\rm{tot}} = \sum_{i=1}^N \langle n_i \rangle.
\end{equation}
We expect $\Delta S$ and $n_{\rm{tot}}$ to approach zero for the initial disordered ground state $\ket{D}$, and to reach their maximal values of $(N+1)/2$ for the target ordered state $\ket{\rm{Z}_2}$ (remind that $N$ is odd). Fig.~\ref{fig6}(d) additionally shows the local Rydberg occupation $\langle n_i \rangle = \langle \psi(\tau) | n_i | \psi(\tau) \rangle$ of the final state $\ket{\psi(\tau)}$, which is expected to exhibit a Z$_2$-staggered pattern. Figs.~\ref{fig6}(b)-(d) clearly reveal that the experimental data are qualitatively consistent with theoretical predictions overall. The two align especially well in the short-time regime but gradually deviate from each other as time progresses.

One aspect worthy of note is the final-state fidelity $F_2(\tau)$, shown in Figs.~\ref{fig6}(b1) and (b2). While the theoretical prediction is close to 1, the experimental results show $F_2 \approx 0.84$ and $F_2 \approx 0.69$ for $N=3$ and $N=7$, respectively. The final state fidelity decreases as $N$ increases. In fact, this issue has also been encountered in adiabatic state preparation, e.g., in Refs.~\cite{Jonathan2023}. The decrease in fidelity mainly stems from an ~8\% measurement error per individual atom \cite{Jonathan2023}. Consequently, for an atom array of length $N$, the measured fidelity will be $0.92^{((N+1)/2)}$, which would approximately be $0.85$ and $0.71$ for $N=3$ and $N=7$, respectively, aligning well with the experimental results shown in Fig.~\ref{fig6}(b).

\begin{figure}[b]
	\begin{center}
		\includegraphics[width=.35 \textwidth]{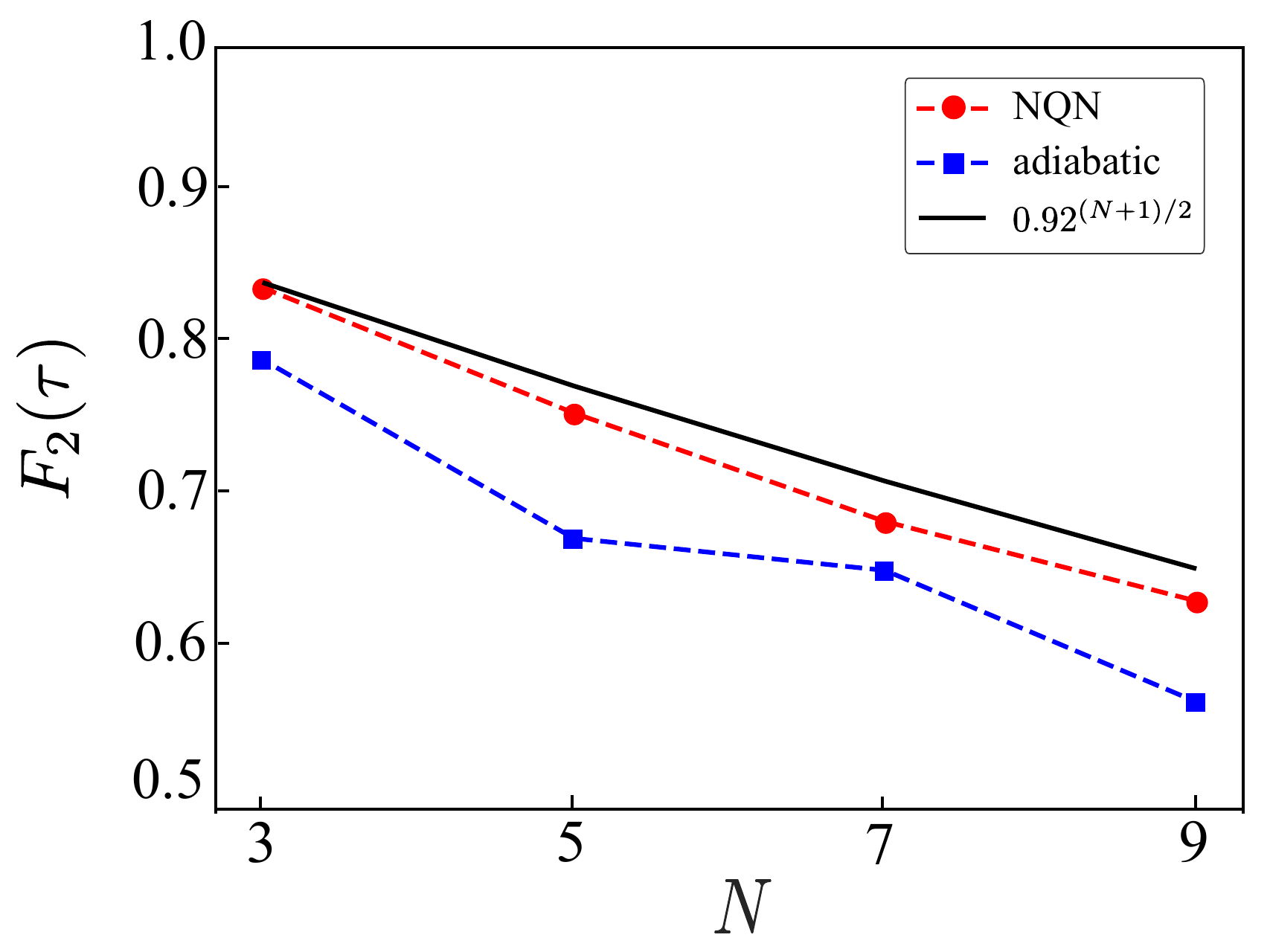}
	\end{center}
		\caption{Final state fidelity $F_2(\tau)$ as a function of particle number $N$. The circles and squares represent the experimental results of the NQN (with $\tau = 1.8\mu s$) and adiabatic (with $\tau = 4\mu s$) schemes, respectively. The solid curve represents the theoretical prediction $0.92^{(N+1)/2}$.}
	\label{fig7}
\end{figure}

In Fig.~\ref{fig7}, we show the final-state fidelity $F_2(\tau)$ obtained by the NQN scheme as a function of $N$ (denoted by circles), with the preparation time $\tau = 1.8\mu\text{s}$ being fixed. We also present the results (marked as squares) of the fully adiabatic scheme by linearly ramping $\Delta$ from $-2.5 (2\pi)$ MHz to $2.5 (2\pi)$ MHz within $4\mu\text{s}$, same as the case in Ref.~\cite{Jonathan2023}. Additionally, the solid curve indicates the theoretical prediction $0.92^{(N+1)/2}$. Both results are shown to be restricted by measurement errors. 
However, compared to the adiabatic scheme, our NQN scheme demonstrates better performance: it exhibits an average improvement of 5\% in fidelity, and is much closer to the optimal measurement results (the solid line).

\section{Conclusion and Outlook} \label{Summary}

To conclude, we have developed an efficient scheme for transitioning from a disordered ground state to various ordered states in a 1D Rydberg atom array. This is achieved through a "non-adiabatic to quasi-adiabatic to non-adiabatic" detuning ramp, known as the NQN scheme. The core principle involves rapidly lifting the system from the ground state to a certain excited state, followed by a quasi-adiabatic transition between the excited state and the target state.
Our calculations demonstrate the capability of the NQN scheme in achieving high-fidelity preparation of $\rm{Z}_2$, $\rm{Z}_3$, and $\rm{Z}_4$ ordered states, with preparation efficiency surpassing traditional adiabatic methods.
Furthermore, the scheme's applicability across systems with varying particle numbers underscores the extensibility of the NQN structure. 
Experimentally, we have implemented the NQN scheme within the permissible operational scope of the Rydberg quantum cloud computer, Aquila.

Looking forward, several concrete steps could further advance this work. Firstly, extending the NQN scheme to higher-order antiferromagnetic ordered states would broaden its applicability. Secondly, applying the NQN scheme to two-dimensional Rydberg atom arrays presents an exciting opportunity. This extension could potentially allow for the preparation of more complex ordered states, such as checkerboard patterns \cite{Scholl2021} or states with topological structures, expanding the toolkit for quantum simulation and computation using Rydberg atoms. 
Furthermore, while the NQN scheme has demonstrated robust performance in systems with a few particles, investigating its scalability for larger systems remains a crucial next step. The key challenge lies in optimizing the time sequence for the Q segment as the system size exceeds the calculation capabilities of traditional numerical algorithms.
Machine learning-based optimization algorithms, such as neural-network quantum states \cite{GuoShuaiFeng2021,CaoJiaHao2023}, could potentially offer valuable solutions to this problem.
Additionally, exploring the integration of the NQN scheme with quantum error correction protocols and quantum error correction devices \cite{Bluvstein2024} could potentially improve its performance in practical quantum computing scenarios.

\begin{acknowledgments}
	We acknowledges support from the NSF of China (Grants No. 12174236, and No. 12222409) and from the fund for the Shanxi 1331 Project.
\end{acknowledgments}

\bibliography{refs}

\end{document}